\documentclass[aps,prb,twocolumn]{revtex4}
\usepackage{graphicx}

\begin{document}

\title{Nontrivial dependence of dielectric stiffness and SHG on \textit{dc} bias in relaxors
and dipole glasses}

 \author{S. A. Prosandeev$^{1}$, I. P. Raevski$^{1}$, A. S. Emelyanov$^{1}$, Eugene
V. Colla$^{2}$, J-L. Dellis$^{3}$, M. El Marssi$^{3}$, S. P.
Kapphan$^{4}$, and L. Jastrabik$^{5}$}
 \affiliation{
$^{1}$\textit{Physics Department, Rostov State University, 344090
Rostov on Don, Russia} \\ $^{2}$\textit{Department of Physics,
University of Illinois at Urbana Champaign, Urbana, Illinois
61801} \\ $^{3}$\textit{LPMC, Universite de Picardie Jules Verne,
33, rue Saint Leu, F-80039 Amiens, France} \\ $^{4}$\textit{FB
Physik,
University of Osnabr\"{u}ck, D-49069 Osnabr\"{u}ck, Germany} \\
$^{5}$\textit{Institute of Physics AS CR, Na Slovance 2, 182 21
Prague 8, Czech Republic} }

\date{\today}

\begin{abstract}
Dielectric permittivity and Second Harmonic Generation (SHG)
studies in the field-cooled mode show a linear dependence of
dielectric stiffness (inverse dielectric permittivity) on
\textit{dc} bias in PMN-PT crystals and SHG intensity in
KTaO$_{3}$:Li at small Li concentrations. We explain this unusual
result in the framework of a theory of transverse,
hydrodynamic-type, instability of local polarization.
\end{abstract}

\maketitle

\section{Introduction}

Locally disordered ferroelectrics, such as $(1-x)$PbMg$_{1 /
3}$Nb$_{2 / 3}$O$_{3} - x$PbTiO$_{3}$ (PMN-PT \cite{1}),
KTaO$_{3}$:Li \textit{etc.}, exhibit unusual properties, which
find wide applications. \cite{2} For example, PMN-PT crystals from
the morphotropic phase boundary (MPB) compositional range possess
an extremely high piezoelectric coefficient, which is important to
transform electric energy to mechanical and back. \cite{3}

The high dielectric permittivity, $\varepsilon$, and its
nonlinearity, i.e. strong voltage dependence of permittivity,
makes disordered ferroelectrics very attractive for applications
in electrically tunable devices, especially in thin-film form. In
these and many other applications (e.g. electrooptic and
electrostriction), disordered ferroelectrics are subjects to high
levels of \textit{dc} biases ($E$). In the framework of Landau
theory, $\varepsilon(E)$~ in cubic crystals is quadratic, at small
fields, and only even terms are allowed. \cite{4}

It was shown \cite{5,6,7} that the giant piezoelectric response in
BaTiO$_{3}$~ and PMN-NT is due to ``\textit{polarization
rotation}'', which means that the electric field turns the
direction of polarization and produces mechanical stresses. It is
important that the potential relief in PMN-PT compositions near
the MPB is nearly flat, which allows large strains at
comparatively small fields.

These materials have some degree of local disorder that is known
to produce polar regions of finite sizes. The rotation of dipole
moments in these regions as well as sideway domain wall movements
(or the domain wall movements in 90$^\circ$~ domains) can
contribute to dielectric permittivity in a special way (which we
will discuss below), especially, when the field-cooled (FC)
procedure is employed (this helps overcoming barriers within the
measuring time).

Another model example considering local dipoles is KTaO$_{3}$:Li
(KLT). Li in KLT is a dipole centre with a large Li-related dipole
moment. \cite{8} Electric field aligns these dipoles and creates
polarization in the direction of the field. \cite{9} If one
changes the direction of the field in the FC mode then the
direction of polarization also changes. That is, the polarization
direction follows the direction of the field that should be if the
Gibbs statistics works. Each of the Li dipoles has some potential
barrier, which should be overcome that produces viscosity. In
ferroelectric phases of PMN-PT, some viscosity can arise also
because of domain wall movements, as well as because of random
fields. The application of the FC procedure allows overcoming this
viscosity.

The ability of an idealized dipole (polarization) to rotate
(without having anisotropy) implies that this dipole has a
transverse instability. Indeed, the application of small fields in
the transverse direction results in a huge (ideally infinite)
response. \cite{10} Random fields, anisotropy, and barriers
suppress this effect at small external fields but some features of
the transverse instability can be seen at larger fields. \cite{11}

In this paper, we present experiments on locally disordered
ferroelectrics, PMN-PT. We will show that dielectric stiffness
(inverse dielectric permittivity), in ferroelectric phases of
PMN-PT, behaves pseudolinearly with the field, above a certain
threshold. This behavior does not follow the Landau theory for
cubic crystals. \cite{4} The suggested in the present paper
theoretical description of this effect takes into account
third-power contributions to the free energy of ferroelectric
phases. The microscopic origin of these contributions is discussed
in detail for a model solid solution, KLT, and we explain
experimental data \cite{9} on the pseudolinear field dependence of
the SHG intensities in KLT at small Li concentrations.

\section{Experiment and results}

The PMN-PT single crystal samples used in this study were
transparent plates cut from a flux grown crystals prepared at the
Physics Research Institute of the Rostov State University.
\cite{12} Large faces of the samples were perpendicular to [100]
direction. The sample with composition PMN-PT35 having dimensions
of about $2 \times 2 \times 1$ mm$^{3}$~ was optically polished
and annealed at 823 K for half an hour in air, in order to
minimize residual stresses. Platinum electrodes were sputtered on
the large faces. Thin (0.05 mm) Pt wires were attached to the
electrodes by silver paste to connect the sample with the contacts
of a Linkam HFS91: TS 600 hot-stage used as a sample holder.
Dielectric measurements were performed at the 2 K/min
heating/cooling rate using a computer-controlled impedance
analyzer Solartron SI 1260. A blocking circuit protected the
impedance analyzer for experiments under high \textit{dc} bias.
Some details of dielectric studies of other PMN-PT crystals have
been described elsewhere. \cite{13,14} Though we focus mainly on
the results obtained in the zero field cooling (ZF) and field
cooling (FC) modes, field heating (FH), and zero field heating
after field cooling protocols were used as well.

\begin{figure}
\resizebox{0.55\textwidth}{!}{\includegraphics{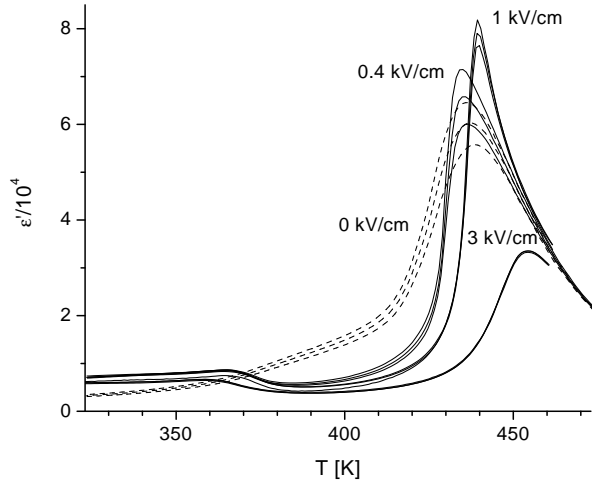}}
\caption{Temperature dependencies of PMN-PT35 dielectric
permittivity measured in the ZF and  FC ($E = 0.4$, 1 and 3 kV/cm)
modes at different frequences:10$^3$, 10$^4$, 10$^5$ Hz (from top
to bottom). }  \label{Fig1}
\end{figure}

The temperature dependences of the complex dielectric permittivity
$\varepsilon(T)$~ measured in the ZF mode for both PMN and
PMN-PT30 single crystals have one diffused frequency-dependent
maximum at about 260 K and 400 K ( at 1 kHz), respectively (see
for details Refs. [\cite{13,14}]). In order to characterize the
PMN-PT35 sample, we show the $\varepsilon(T)$~ curves measured in
the ZF mode at different frequencies (Fig.1). Two anomalies on the
$\varepsilon(T)$~ curve are in good agreement with those found in
Refs. [\cite{13,15,16,17,18}]: a diffused maximum corresponding to
the cubic-to-tetragonal phase transition and an inflexion at a
lower temperature, $T_{1}$, corresponding to the MPB [the
transition between the tetragonal and rhombohedral (or monoclinic,
according to different data) \cite{13,15,16,17}] phases. The
temperature $T_{m}$~ of the $\varepsilon(T)$~ maximum (435 K at 1
KHz) agrees well with the data obtained for PMN-PT crystals of
similar compositions, \cite{16,18} The temperature $T_{1}$~ ($
\approx $365-375 K at 1 KHz) is comparable with the data for
flux-grown crystals with similar $T_{m}$~ values, \cite{15} but is
substantially (60-100 K) higher than $T_{1}$~ values, reported for
PMN-PT crystals with similar $T_{m}$~ values grown by Bridgeman
method. \cite{17,18} Such discrepancy seems to be caused by
differences in crystal's preparation technique and its origin
needs additional studies.

\begin{figure}
\resizebox{0.45\textwidth}{!}{\includegraphics{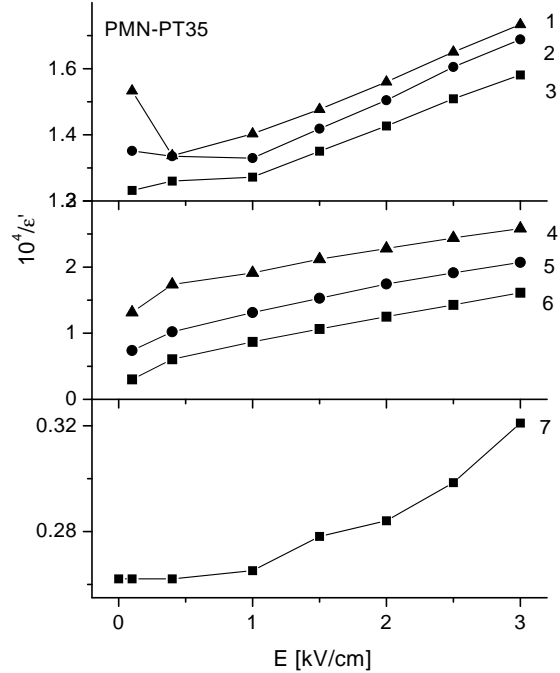}}
\caption{Dielectric stiffness vs. the dc bias for [100] PMN-PT35
single crystals measured at different temperatures under FC
protocol at 10 kHz: Rhombohedral (monoclinic) phase, 323 K (1),
339 K (2), 353 K (3); Tetragonal phase, 383 K (4), 413 K (5), 423
K (6); Cubic phase, 459 K (7).Solid lines are guides to the eye. }
\label{Fig2}
\end{figure}

In ZF mode, a pronounced dielectric dispersion is observed below
$T_{m}$~ in PMN-PT35 and $T_{m}$~ increases with the frequency
$f$~ of the measuring electric field. The $T_{m}(f)$~ dependence
obeys the Vogel-Fulcher relation. \cite{19} While the value of the
attempt frequency $f_0 =5 \times 10^{11}$~ Hz is typical for
relaxors, \cite{19} the Vogel-Fulcher temperature  $T_0 = 430\,
\mathrm{K}$~ is only slightly lower than $T_{m}$. Though
dielectric dispersion decreases below $T_{0}$, it remains
considerable in the ZF mode (Fig.1). In the FC mode, the
dispersion in the tetragonal phase diminishes but, at low $E$'s,
remains practically unchanged, close to $T_m$. Such behavior is
similar to that observed in disordered PST and PSN-based ceramics
and crystals \cite{19,20,21} and seems to be due to the presence
of the spontaneous (thermally driven) transition from relaxor to a
mixed (ferroelectric/relaxor) phase containing both the nanoscale
polar regions and macroscopic ferroelectric domains of tetragonal
symmetry. Both the $T_m$~ and $T_1$~ exhibit thermal hysteresis,
in agreement with first order type reported for these phase
transitions. \cite{16,17,18} The degree of the $\varepsilon(T)$~
maximum diffusion evaluated using approach \cite{22} is about 20
K, in good agreement with the data for PMN-PT31 crystals and
PMN-PT35 ceramics. \cite{22}

\begin{figure}
\resizebox{0.45\textwidth}{!}{\includegraphics{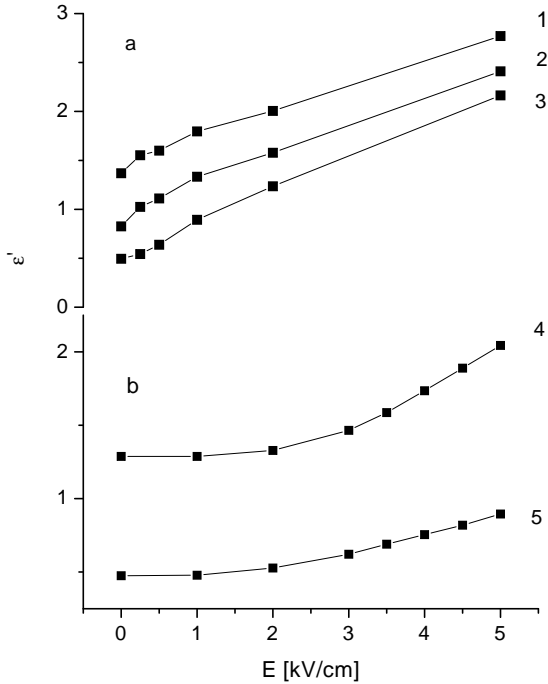}}
\caption{Dielectric stiffness vs the dc bias for [100] PMN-PT30
and [100] PMN single crystals measured at different temperatures
under FC protocol at 1 kHz: 350 K (1), 360 K (2), 370 K (3), 200 K
(4), 250 K (5). Solid lines are guides to the eye. } \label{Fig3}
\end{figure}

In the FC mode, the application of even a relatively small
\textit{dc} bias leads to substantial changes of $\varepsilon
(T)$. Permittivity in the tetragonal phase decreases with the bias
while, in the rhombohedral phase, it increases initially, but, at
higher biases, begins to decrease (Fig. 1). As a result, the
inflexion transforms into a maximum at $T_{1} \approx $ 363 K, in
good agreement with the data for the [001]-oriented PMN-PT
crystals from the $0.31<x<0.35$~ compositional range. \cite{17,18}

As the \textit{dc} bias increases, the main \textit{$\varepsilon
$}($T)$ maximum, at first, becomes sharper and higher but then
lowers, diffuses and shifts to higher temperatures (Fig. 1). These
changes are in good agreement with the data for the [001]-oriented
PMN-PT crystals from the $0.31< x <0.35$~ compositional range
\cite{17,18} as well as with the data on disordered PbSc$_{1 /
2}$Ta$_{1 / 2}$O$_{3}$~ ceramics. \cite{23} The frequency
dispersion of $\varepsilon$~ and $T_{m}$~ reduces dramatically for
biases exceeding (0.7-1) kV/cm (Fig. 1). Thus, at low $E$, the
PMN-PT35 crystal exhibits a relaxor-like behavior, while, at high
$E$'s, its properties are similar to ordinary ferroelectrics. The
details of the dependence of $T_{m}$~ and $T_{1}$~ on $E$~ will be
a subject of a separate publication; here we focus on the field
dependences of dielectric stiffness.

The main new result, which we want to discuss in the present
paper, is the pseudolinear field dependence of dielectric
stiffness in the ferroelectric phases of PMN-PT35 above a certain
threshold (Fig. 2). Such behavior is not unique. For comparison,
Fig. 3 presents dielectric stiffness \textit{vs.} $E$~ for [100]
PMN-PT30 and PMN crystals. It is seen that, the pseudolinear
dependence is fulfilled for these compositions as well, but above
a larger threshold.

\section{Theory}

\subsection{Hydrodynamic model}

At first, we want to discuss experimental data obtained for model
compounds, KTaO$_{3}$:Li. This case has been studied carefully on
the basis of first-principles computations (see \cite{8} and
references therein) and experiment. \cite{9} It was obtained that
Li's are off-center and have a large dipole moment. Li-related
dipoles (further, Li-dipoles) can occupy 6 possible positions with
equal energy. Hence, each dipole can rotate (FC protocol is
assumed). Let a dipole occupy a position along $z$. The
application of field $\delta E$~ perpendicular to $z$~ rotates the
dipole. The change of polarization can be expressed as:

\begin{equation}
\label{eq1} \delta P_\perp = \chi_\perp \delta E = n\mu \delta E /
E
\end{equation}
where $n$~ is the Li concentration, $\mu$~ the dipole moment, $E$~
the magnitude of the field along axis $z$, and $\chi_\perp $ the
transverse susceptibility: $\chi_\perp $ diverges in this
simplified approach, \cite{9} but taking into account the random
fields and anisotropy of the potential relief stabilizes the
system. \cite{11}

\begin{figure}
\resizebox{0.45\textwidth}{!}{\includegraphics{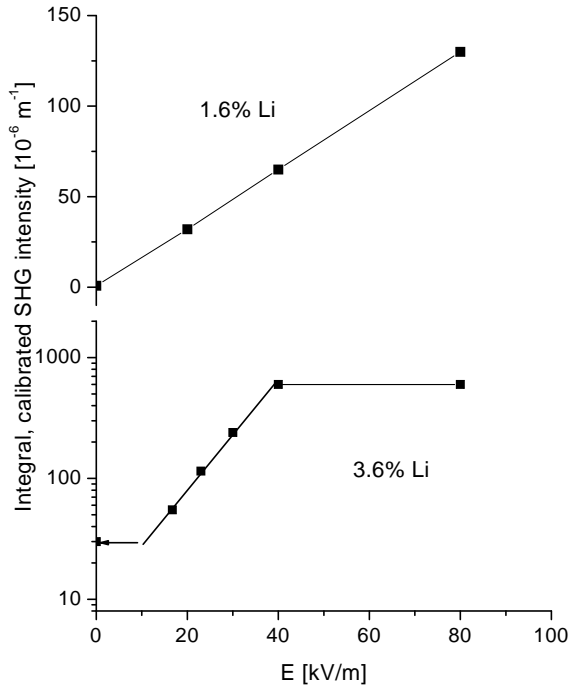}}
\caption{The field dependence of the FC SHG intensity in KLT
\cite{9}. Solid lines are guides to the eye. } \label{Fig4}
\end{figure}

The change of the polarization magnitude in the field can be
described by the Landau theory based on the soft mode concept,
which we will discuss below. The free energy necessary for the
description of the transverse fluctuations has the form \cite{24}

\begin{equation}
\label{eq2} \delta F = \frac{1}{2}\int {\left[ {\chi_\perp ^{ - 1}
\delta P_ \perp ^2 + c\left( {\nabla P_ \perp } \right)^2}
\right]} dV
\end{equation}

\noindent where $c$~ is constant. One can obtain the longitudinal
susceptibility from relations between the transverse and
longitudinal changes of polarization \cite{9}:

\begin{eqnarray}
\label{eq3} &&\chi _{\parallel} = \frac{\delta P_{\parallel }
}{\delta E} = \frac{1}{2P}\frac{d}{dE}\left\langle {\left( {\delta
P_ \perp } \right)^2} \right\rangle = \frac{nk_B T}{8\pi \left(
{\left\langle \mu \right\rangle c} \right)^{3 / 2}E^{1 / 2}}
\nonumber \\ && E > 0
\end{eqnarray}

Notice that this susceptibility is due to transverse fluctuations
of polarization. Normally, it should be added to the Langevin
susceptibility that will be discussed below. The integral of this
equation gives \cite{25}

\begin{eqnarray}
\label{eq4} && \left( {P - P_0 } \right)^2 = aE \nonumber \\
&& P > P_0 \nonumber \\
&& E> 0
\end{eqnarray}

\noindent where $a = \left( {nk_B T} \right)^2 / 16\pi ^2\left(
{\left\langle \mu \right\rangle c} \right)^3$ and $P_0 $ is the FC
polarization magnitude at small $E$ (remnant polarization). Random
fields modify this equation of state. \cite{11} It was obtained
that equation (\ref{eq4}) is true for the field magnitude, which
is larger than the random-field magnitude. It is important to say
that equation (\ref{eq4}) corresponds to a cubic contribution to
free energy \cite{10}:

\begin{eqnarray}
\label{eq5} && \Delta F = A\left( {P - P_0 } \right)^3 - EP
\nonumber
\\
&& P> P_0
\end{eqnarray}
where $A = a/3$.

Notice, that the same conclusions can be got under assumption of
domain wall contributions to dielectric permittivity.
Experimentally, for dielectrics, some violations of Landau
expansion over the even terms (valid for the paraelectric phase
only) were discussed for PLZT by J-L. Delis \cite{26}; the author
\cite{26} connected this with domain wall contributions. The
evidence of equation (\ref{eq4}) in dielectrics was found in
experiments on SHG \cite{9} (in magnetic systems it was discussed
in \cite{10,25}).

The SHG intensity, $S$, is proportional to the average square of polarization:

\begin{eqnarray}
\label{eq6} && S\sim \left( {P_0 + \sqrt {aE} } \right)^2
\nonumber
\\
&& E> 0
\end{eqnarray}

It gives the linear dependence of the FC SHG intensity on $E$~ at
$aE \gg P_0^2 $. This finding is in agreement with the SHG
experiment performed for KLT at Li concentration 1.6{\%} (KLT
1.6{\%}), in the FC mode (see Fig. 4).

\subsection{Pseudolinear field dependence of dielectric
stiffness}

The above consideration is true if only the lattice (soft mode)
contribution to polarization is comparatively negligible.
Experiments on KLT \cite{27} showed that the lattice (soft mode)
contribution to SHG intensity starts at much larger fields than
those used in experiment. \cite{9} In some other cases, the
contribution of the soft mode may be comparable with the
contribution of local dipoles, polar regions or domain walls. We
think that this is the case in PMN-PT35. Due to low symmetry of
ferroelectric phases (we are talking about the phases obtained in
the FC mode at $E \to 0)$, free energy includes an odd term:

\begin{eqnarray}
\label{eq7} && F = \frac{1}{2}\alpha P^2 + \frac{1}{3}AP^3 +
\frac{1}{4}bP^4 - EP + ... \nonumber \\ && P > P_0 = - \alpha / A
\end{eqnarray}
The equilibrium condition gives

\begin{eqnarray}
\label{eq8} && \alpha P + AP^2 + bP^3 + ... - E = 0 \nonumber
\\ && P> P_0
\end{eqnarray}
Now one can obtain that

\begin{eqnarray}
\label{eq9} && \chi ^{ - 1} = \alpha + 2AP + 3bP^2 + ...\nonumber
\\ && P> P_0
\end{eqnarray}
At small fields, $P = - \alpha / A + \left| \alpha \right|^{ -
1}E$~ that leads to the observed linear field dependence of
inverse susceptibility \cite{28}:

\begin{eqnarray}
\label{eq10} && \chi ^{ - 1} = - \alpha + 3bP_0^2 + \frac{2A +
6bP_0 }{\left| \alpha \right|}E + ... \nonumber \\ && E > 0
\end{eqnarray}
At large fields, one needs to add even terms, and the dependence
becomes quadratic and, then, $E^{2/3}$~ as it should be in the
Landau theory developed for cubic crystals. The interval of the
fields, where (\ref{eq10}) is true, depends on the coefficients in
expansion (\ref{eq7}). One can expect this effect close to the
phase transition temperature, and at large $A$, which is possible
when the transverse spatial fluctuations of polarization are
large. Both conditions are satisfied at MPB.

\subsection{Influence of random fields on the FC dependence of
dielectric stiffness in PMN}

The influence of random fields on hydrodynamic fluctuations of
polarization results in the insensitivity of dielectric
permittivity to external fields at small fields. \cite{11} Here we
will discuss another aspect of this problem. Experiment shows
that, in the ferroelectric phases of PMN-PT35, dielectric
stiffness pseudolinearly depends on the field in a wide interval
of the field. Only at very small fields, where, perhaps, the time
of measuring is not enough, there are some deviations from the
pseudolinear dependence (Fig. 2). Experimental results obtained
for pure PMN \cite{14} show a more pronounced influence of the
random fields on dielectric stiffness. It was found that, in the
[111] direction, dielectric stiffness of PMN has a bump while, in
the [001] direction, dielectric stiffness decreases with the field
smoothly. We suggest considering this difference as a result of
anisotropy of free energy. We take into account that there is a
diffuse first order phase transition in the [111] field. \cite{14}

\begin{figure}
\resizebox{0.55\textwidth}{!}{\includegraphics{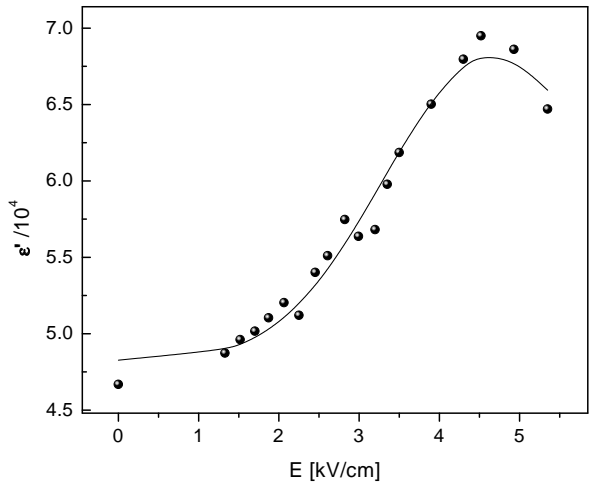}}
\caption{The fit of theory (solid line) to experiment (points) for
PMN [111] obtained in the FC protocol at T=250 K and frequency
=100 Hz. } \label{Fig5}
\end{figure}

We consider the free energy in [111] direction corresponding to the
first-order phase transition between the cubic and rhombohedral phases:

\begin{equation}
\label{eq11}
F\left( {e,T_c } \right) = \frac{1}{2}\alpha P^2 - \frac{1}{4}\beta P^4 +
\frac{1}{6}\gamma P^6 - \left( {E + e} \right)P
\end{equation}
where $\alpha = a\left( {T - T_c } \right)$ and $e$ is random
field. From the equilibrium condition, one gets:

\begin{equation}
\label{eq12}
\alpha P - \beta P^3 + \gamma P^5 = E + e
\end{equation}
Dielectric permittivity can be found by differentiating
(\ref{eq12}):

\begin{equation}
\label{eq13}
\varepsilon '\left( {e,T_c ,E} \right) = 1 + \frac{1}{\varepsilon _0
}\frac{dP}{dE} = 1 + \frac{1}{\varepsilon _0 }\frac{1}{\alpha - 3\beta P^2 +
5\gamma P^4}
\end{equation}
In the case $e$ = 0, polarization experiences a jump at some field
E$_{0}$, and dielectric permittivity diverges at this field. Due
to the random fields and distribution of $T_{c}$'s, this jump
diffuses. \cite{29,30} This can be described by introducing a
distribution function $f(e$,$T_{c})$ (see the way how to determine
the distribution function from experiment in Ref. 30):

\begin{equation}
\label{eq14} \varepsilon _{av} (E) = \int {\varepsilon (e,T_c,E)
f(e,T_c)dedT_c}
\end{equation}
In order to simplify this complex situation, one can model
inhomogeneous matter by introducing a Gauss type distribution of
$E_{0}$:

\begin{eqnarray}
\label{eq15} && P(E) = P_1(E) + \nonumber \\ && +P_2(E)\int
{\theta (E_0 - E) e^{[-(E_0 - E_1)/x_0]^2}dE_0 }
\end{eqnarray}
where $E_{1}$ is the position of the dielectric permittivity
maximum, \textit{$\theta $(x)} is a step function, $x_{0}$ is the
width of the distribution function, $P_{1}(E)$ and $P_{2}(E)$ are
continuous functions. Dielectric permittivity can be got by
differentiating (\ref{eq15}):

\begin{eqnarray}
\label{eq16} && \varepsilon(E) = \alpha_1(E) - \nonumber \\ &&
-\alpha_2 (E) \textrm{erf}[(E - E_1)/x_0] + \nonumber
\\ && + P_2 \left( E \right) e^{-[(E - E_1)/x_0]^2}
\end{eqnarray}
here erf$(x)$~ is the probability integral, $\alpha _1 $ and
$\alpha _2 $ can be expressed over the derivatives of $P_{1}(E)$
and $P_{2}(E)$.

Fig. 5 shows the result of the fitting of our expression to
experiment. The distribution function width was found at 1.97
kV/cm, and $E_{c}$ = 4.27 kV/cm. For the sake of simplicity, we
took $\alpha _1 $ and $\alpha _2 $ as constants, and $P_{2}$ was
proportional to $E$. The fit shows that the nature of the bump in
the dependence of dielectric permittivity on the [111] field can
be the diffuse anomaly due to the first order phase transition
between the cubic and rhombohedral phases.

\section{Discussion}

The performed experiment reveals that dielectric stiffness in the
ferroelectric phases of PMN-PT from the MPB range behaves linearly
with $E$. By integrating (\ref{eq10}) we obtained that the
dependence of polarization on $E$ is logarithmic. This dependence
includes both even and odd terms although, in the paraelectric
phase only even terms are expected. In ferroelectric phases, the
odd terms are allowed due to extrinsic contributions. \cite{26}
Notice that the paraelectric phase can be a result of ZF
procedure. Cooling the same sample at finite $E$ can result in the
appearance of remnant polarization. Our experiments have shown
that the pseudolinear field dependence of dielectric stiffness is
well observed at temperatures near the MPB (where an intermediate
monoclinic phase can appear stimulating the polarization rotation
\cite{5,7}), but, in the paraelectric phase, the data are not that
definite. Thus the important feature for the observation of the
linearization of dielectric stiffness \textit{vs} $E$ is a flat
enough potential relief, which is indeed the case in PMN-PT
compositions from the MPB range. Note, that in the PMN-PT30
crystal, which is outside of the morphotropic phase boundary
region and especially in pure PMN, the pseudolinear portion of the
1/\textit{$\varepsilon $(E)} dependence is observed at larger
fields than in PMN-PT35. This is consistent with theory, \cite{11}
which showed that the onset of the unusual power law in the
hydrodynamic model is at the fields, which are larger then the
random field magnitude. In that study only the cubic term was
considered in the free energy. In the present study we have added
the normal quadratic term responsible for the polarization of
lattice (soft-mode contribution). We have shown that these two
terms together result in the pseudolinear field-dependence of
nonlinear permittivity.

Our finding correlates with the recent observation of the linear
field dependence of dielectric stiffness in thin films of PMN
where the energy barriers between metastable states of polar
clusters appear smaller than in bulk and ceramics. \cite{31}
Approximately linear dependences of dielectric stiffness on
\textit{dc} bias were observed in disordered PST ceramics near the
border of stability of relaxor \textit{vs} ferroelectric phase.
\cite{23}

At small $P_0 $, the expression obtained for dilute KLT provides
the uncommon equation of state, $P^2\sim E$, which resembles that
obtained for some magnets. \cite{25} This behavior has been
evidenced in experiments on SHG in KTaO$_{3}$:Li for the Li
concentrations 1.6{\%} \cite{8} (Fig. 4), which, in our opinion,
reflects the fact of strong transversal spatial fluctuations of
polarization. If the ZF procedure were employed then one would not
expect such an effect at low T due to the nonergodicity in this
system arising because of potential barriers (viscosity). In such
experiments only the ordinary, Langevin type, response of the
dipoles should be seen. However, if, due to different reasons, a
tilt of the Li dipoles is prevented by internal random electric
fields or stresses then the fluctuations become weak and the
corresponding contribution to the susceptibility vanishes. For
example, for the concentrations larger than 2{\%}, the SHG
intensity behaves with the field exponentially, $\sim \exp \left[
{\left( {\mu E - w} \right) / k_B T} \right]$ (Fig. 4, notice the
logarithmic scale). It can be connected with the fact that the Li
dipoles are merged into large clusters or even domains at these
concentrations. A strong Li -- Li interaction prevents their
reorientation. It implies that there is energy, $w$, which is
necessary to reorient dipoles. The electric field applied
decreases this energy linearly. Such dependence can be seen even
for low Li concentrations if the sample is strained. There can be
also some differences due to different protocol in the FC
procedure (for instance, due to different annealing temperatures).

Finally, we have observed a pseudolinear field dependence of
dielectric stiffness in [100] PMN, PMN-PT35 and PMN-PT30 single
crystals in the FC mode. We have explained this dependence by
polarization fluctuations (rotations) and domain wall movements.
\cite{26} Within the same approach we managed to explain the
experimental data \cite{9} on the pseudolinear dependence of the
SHG intensities on $E$ in KLT at small Li concentrations.

I. P. R. appreciates a financial support from Scientific Research
Fellowship and hospitality of LPMC, Universite de Picardie Jules
Verne. S. A. P. is grateful for grants ru.01.01.037 (``Russian
Universities''), 04-02-16103 (RFBR), NIST, Maryland, for
hospitality, and Sergey Vakhrushev for discussing the data on the
first order phase transition in PMN. This study is partially
supported by LN00A015 of the MSMT CR.


\begin{thebibliography}{999}
\bibitem{1}In some cases, we will show $x$, expressed in molar percents,
after PMN-PT.
\end{thebibliography}
\end{document}